\documentclass[modern]{aastex62}

\usepackage{amsmath}
\usepackage{amssymb}
\usepackage{rotating}

\newcommand\tE{t_{\rm E}}
\newcommand\thetaE{\theta_{\rm E}}

\newcommand\murel{\mu_{\rm rel}}
\newcommand\pirel{\pi_{\rm rel}}
\newcommand{\ra}[4]{${#1}^{\rm h}{#2}^{\rm m}{#3}\fs{#4}$}
\newcommand{\dec}[4]{${#1}\arcdeg{#2}\arcmin{#3}\farcs{#4}$}

\newcommand\Fs{F_{\rm s}}
\newcommand\Fb{F_{\rm b}}

\received{January 1, 2018}
\revised{January 1, 2018}
\accepted{January 1, 2018}

\shorttitle{A rogue planet in the microlensing event OGLE-2019-BLG-0551}
\shortauthors{Mr\'oz et al.}

\begin{document}

\title{A free-floating or wide-orbit planet in the microlensing event OGLE-2019-BLG-0551}

\correspondingauthor{Przemek Mr\'oz}
\email{pmroz@astro.caltech.edu}

\author[0000-0001-7016-1692]{Przemek Mr\'oz}
\affil{Division of Physics, Mathematics, and Astronomy, California Institute of Technology, Pasadena, CA 91125, USA}
\affil{Astronomical Observatory, University of Warsaw, Al. Ujazdowskie 4, 00-478 Warszawa, Poland}

\author[0000-0002-9245-6368]{Rados\l{}aw Poleski}
\affil{Astronomical Observatory, University of Warsaw, Al. Ujazdowskie 4, 00-478 Warszawa, Poland}

\author{Cheongho Han}
\affil{Department of Physics, Chungbuk National University, Cheongju 28644, Republic of Korea}

\author[0000-0001-5207-5619]{Andrzej Udalski}
\affil{Astronomical Observatory, University of Warsaw, Al. Ujazdowskie 4, 00-478 Warszawa, Poland}

\author{Andrew Gould}
\affil{Max Planck Institute for Astronomy, K\"{o}nigstuhl 17, D-69117 Heidelberg, Germany}
\affil{Department of Astronomy, Ohio State University, 140 W. 18th Ave., Columbus, OH 43210, USA}

\nocollaboration 

\author[0000-0002-0548-8995]{Micha\l{} K. Szyma\'nski}
\affil{Astronomical Observatory, University of Warsaw, Al. Ujazdowskie 4, 00-478 Warszawa, Poland}

\author[0000-0002-7777-0842]{Igor Soszy\'nski}
\affil{Astronomical Observatory, University of Warsaw, Al. Ujazdowskie 4, 00-478 Warszawa, Poland}

\author[0000-0002-2339-5899]{Pawe\l{} Pietrukowicz}
\affil{Astronomical Observatory, University of Warsaw, Al. Ujazdowskie 4, 00-478 Warszawa, Poland}

\author[0000-0003-4084-880X]{Szymon Koz\l{}owski}
\affil{Astronomical Observatory, University of Warsaw, Al. Ujazdowskie 4, 00-478 Warszawa, Poland}

\author[0000-0002-2335-1730]{Jan Skowron}
\affil{Astronomical Observatory, University of Warsaw, Al. Ujazdowskie 4, 00-478 Warszawa, Poland}

\author[0000-0001-6364-408X]{Krzysztof Ulaczyk}
\affil{Department of Physics, University of Warwick, Coventry CV4 7 AL, UK}
\affil{Astronomical Observatory, University of Warsaw, Al. Ujazdowskie 4, 00-478 Warszawa, Poland}

\author[0000-0002-1650-1518]{Mariusz Gromadzki}
\affil{Astronomical Observatory, University of Warsaw, Al. Ujazdowskie 4, 00-478 Warszawa, Poland}

\author[0000-0002-9326-9329]{Krzysztof Rybicki}
\affil{Astronomical Observatory, University of Warsaw, Al. Ujazdowskie 4, 00-478 Warszawa, Poland}

\author[0000-0002-6212-7221]{Patryk Iwanek}
\affil{Astronomical Observatory, University of Warsaw, Al. Ujazdowskie 4, 00-478 Warszawa, Poland}

\author[0000-0002-3051-274X]{Marcin Wrona}
\affil{Astronomical Observatory, University of Warsaw, Al. Ujazdowskie 4, 00-478 Warszawa, Poland}

\collaboration{(OGLE Collaboration)}
\noaffiliation

\author{Michael D. Albrow}
\affil{University of Canterbury, Department of Physics and Astronomy,
Private Bag 4800, Christchurch 8020, New Zealand}

\author{Sun-Ju Chung}
\affil{Korea Astronomy and Space Science Institute, Daejon 34055,
Republic of Korea}
\affil{University of Science and Technology, Korea (UST) Gajeong-ro,
Yuseong-gu,  Daejeon 34113, Republic of Korea}

\author{Kyu-Ha Hwang}
\affil{Korea Astronomy and Space Science Institute, Daejon 34055,
Republic of Korea}

\author{Yoon-Hyun Ryu}
\affil{Korea Astronomy and Space Science Institute, Daejon 34055,
Republic of Korea}

\author{Youn Kil Jung}
\affil{Korea Astronomy and Space Science Institute, Daejon 34055,
Republic of Korea}

\author{In-Gu Shin}
\affil{Korea Astronomy and Space Science Institute, Daejon 34055,
Republic of Korea}

\author{Yossi Shvartzvald}
\affil{Department of Particle Physics and Astrophysics, Weizmann Institute of Science, Rehovot 76100, Israel}

\author{Jennifer C. Yee}
\affil{Center for Astrophysics $|$ Harvard \& Smithsonian, 60 Garden St., Cambridge, MA 02138, USA}

\author{Weicheng Zang}
\affil{Department of Astronomy and Tsinghua Centre for Astrophysics, Tsinghua University, Beijing 100084, China}

\author{Sang-Mok Cha}
\affil{Korea Astronomy and Space Science Institute, Daejon 34055,
Republic of Korea}
\affil{School of Space Research, Kyung Hee University, Yongin, Kyeonggi 17104, Republic of Korea}

\author{Dong-Jin Kim}
\affil{Korea Astronomy and Space Science Institute, Daejon 34055,
Republic of Korea}

\author{Hyoun-Woo Kim}
\affil{Korea Astronomy and Space Science Institute, Daejon 34055,
Republic of Korea}

\author{Seung-Lee Kim}
\affil{Korea Astronomy and Space Science Institute, Daejon 34055,
Republic of Korea}
\affil{University of Science and Technology, Korea (UST) Gajeong-ro,
Yuseong-gu,  Daejeon 34113, Republic of Korea}

\author{Chung-Uk Lee}
\affil{Korea Astronomy and Space Science Institute, Daejon 34055,
Republic of Korea}
\affil{University of Science and Technology, Korea (UST) Gajeong-ro,
Yuseong-gu,  Daejeon 34113, Republic of Korea}

\author{Dong-Joo Lee}
\affil{Korea Astronomy and Space Science Institute, Daejon 34055,
Republic of Korea}

\author{Yongseok Lee}
\affil{Korea Astronomy and Space Science Institute, Daejon 34055,
Republic of Korea}
\affil{School of Space Research, Kyung Hee University, Yongin, Kyeonggi 17104, Republic of Korea}

\author{Byeong-Gon Park}
\affil{Korea Astronomy and Space Science Institute, Daejon 34055,
Republic of Korea}
\affil{University of Science and Technology, Korea (UST) Gajeong-ro,
Yuseong-gu,  Daejeon 34113, Republic of Korea}

\author{Richard W. Pogge}
\affil{Department of Astronomy, Ohio State University, 140 W. 18th Ave., Columbus, OH 43210, USA}

\collaboration{(KMT Collaboration)}
\noaffiliation

\begin{abstract}
High-cadence observations of the Galactic bulge by the microlensing surveys led to the discovery of a handful of extremely short-timescale microlensing events that can be attributed to free-floating or wide-orbit planets. Here, we report the discovery of another strong free-floating planet candidate, which was found from the analysis of the gravitational microlensing event OGLE-2019-BLG-0551. The light curve of the event is characterized by a very short duration ($\lesssim 3$\,days) and a very small amplitude ($\lesssim 0.1$\,mag). From modeling of the light curve, we find that the Einstein timescale, $\tE=0.381 \pm 0.017$\,day, is much shorter, and the angular Einstein radius, $\thetaE=4.35 \pm 0.34$\,$\mu$as, is much smaller than those of typical lensing events produced by stellar-mass lenses ($\tE\sim 20$\,days, $\thetaE\sim 0.3$\,mas), indicating that the lens is very likely to be a planetary-mass object. We conduct an extensive search for possible signatures of a companion star in the light curve of the event, finding no significant evidence for the putative host star. For the first time, we also demonstrate that the angular Einstein radius of the lens does not depend on blending in the low-magnification events with strong finite source effects.

\end{abstract}


\section{Introduction} \label{sec:intro}

Gravitational microlensing is the only technique that allows us to detect low-mass rogue (free-floating) planets, that is, planetary-mass objects that are not gravitationally tethered to any star. Microlensing events caused by free-floating planets are characterized by small angular Einstein radii ($\thetaE\lesssim 10\,\mu$as) and extremely short timescales ($\tE \lesssim 1\,$day), rendering them difficult to detect and requiring frequent photometric observations (with a frequency of $\sim 10$ per night per site or higher). 

\citet{mroz2017} created an unbiased sample of 2617 microlensing events observed at a high cadence by the Optical Gravitational Lensing Experiment (OGLE) survey \citep{udalski2015} and discovered an excess of six short-timescale ($\tE\lesssim 0.5$\,day) microlensing events that can be attributed to Earth- and Neptune-mass objects. Timescales of microlensing events depend on their angular Einstein radius $\thetaE$ and relative lens-source proper motion $\murel$:
\begin{equation}
\tE = \frac{\thetaE}{\murel},
\end{equation}
making it possible that an unusually high proper motion results in very short timescales. Here $\thetaE=\sqrt{\kappa M \pirel}$, where $\kappa=8.144\,\mathrm{mas}\,M_{\odot}^{-1}$, $M$ is the lens mass, and $\pirel$ is relative lens-source parallax. 

Luckily, it was possible to measure $\thetaE$ in a handful of short-timescale events \citep[e.g., OGLE-2012-BLG-1323, $\tE=0.155 \pm 0.005$\,day, $\thetaE=2.37 \pm 0.10$\,$\mu$as; OGLE-2016-BLG-1540, $\tE=0.320 \pm 0.003$\,day, $\thetaE=9.2 \pm 0.5$\,$\mu$as;][]{mroz2018,mroz2019} thanks to finite source effects and, therefore, to confirm that their short timescales result from small Einstein radii of the lensing objects. Finite source effects are observed whenever the limb of the source passes over/near the position of the lens \citep{gould1994,witt1994,nemiroff1994}. We expect that microlensing events due to planetary-mass objects should exhibit strong finite source effects because as the mass of the lens gets smaller, the angular Einstein radius becomes comparable to angular radii of the source stars \citep{bennett1996}. In particular, if the angular size of the star being lensed is much larger than the angular Einstein radius, the maximal magnification is suppressed \citep{witt1994,gould1997},
\begin{equation}
A_{\rm max} \rightarrow 1+2\left(\frac{\thetaE}{\theta_*}\right)^2 \ \mathrm{if}\ \thetaE \ll \theta_*,
\label{eq:amax}
\end{equation}
where $\theta_*$ is the angular radius of the source. This is the case in microlensing events OGLE-2012-BLG-1323 and OGLE-2016-BLG-1540, both of which occurred on large giant sources.

Unfortunately, microlensing observations alone are usually not able to distinguish between free-floating and wide-orbit planets \citep{han2003,han2005}. For a few published microlensing events probably due to free-floating planets, we may provide only lower limits on the projected separation of the putative host stars, which are on the order of 5--10\,au, depending on their distance \citep{mroz2018,mroz2019}.

Owing to large orbital separations and long orbital periods, detecting and measuring the frequency of (bound) wide-orbit planets is challenging. Exoplanet direct imaging surveys enable discovering the most massive giants planets, with the estimated occurrence rate of $11^{+11}_{-5}\%$ (between $1-20\,M_{\rm Jup}$ and $5-5000$\,au, around nearby stars; \citealt{baron2019}) or $9^{+5}_{-4}\%$ (between $5-13\,M_{\rm Jup}$ and $10-100$\,au, around $M>1.5\,M_{\odot}$ stars; \citealt{nielsen2019}). A few wide-orbit low-mass planets were detected in microlensing events OGLE-2008-BLG-092 \citep[$s=5.3$;][]{poleski2014}, OGLE-2011-BLG-0173 \citep[$s=4.6$;][]{poleski2018b}, and KMT-2016-BLG-1107 \citep[$s=3.0$;][]{hwang2019}, where $s$ is the projected separation in Einstein radius units. These values correspond to physical separations from 7 to 15\,au and parameters of OGLE-2008-BLG-092Lb are at the edge of current limits of detecting putative hosts of free-floating planets. Recently, the DSHARP project \citep{andrews2018} published deep, high-resolution images of 20 nearby protoplanetary disks, which show annular substructures that are believed to result from planet--disk interactions \citep{huang2018,zhang2018} (although it should be noted that other mechanisms, such as pebble growth near the snowlines \citep{zhang2015,okuzumi2016} or magneto-rotational instability \citep{flock2015} may also produce gaps and rings in protoplanetary disks). Using DSHARP observations, \citet{zhang2018} inferred that about 50\% of analyzed stars host a Neptune- to Jupiter-mass planet beyond 10\,au.

If the projected separation between a planet and its host star is much larger than the Einstein radius of the system (i.e., $s\gg 1$), the microlensing light curves will look like that of a single object, unless the trajectory of the source happens to pass near both components. In that case, we expect to detect a second low-amplitude brightening in the event light curve due to the host star well before or after the main event. It is also possible to identify wide-separation planetary events from the signature of the planetary caustic near the peak of the light curves. 
For example, \citet{han2019a} conducted a systematic search for short-timescale microlensing events exhibiting finite source effects. A detailed modeling of one of these events, OGLE-2016-BLG-1227, revealed low-amplitude ($\sim 0.03\,$mag) residuals from a single-lens light curve, most likely due to a low-mass host star \citep{han2020}.

In principle, one may distinguish between wide-orbit and free-floating planets from high-resolution images taken well after the event, when the lens and source separate. This is challenging in case of detected free-floating planet candidates because the sources are bright giant stars, and so the high contrast renders detection of putative stellar companions difficult. However, this will become routine with the advent of adaptive optics (AO) on 30\,m class telescopes \citep[e.g.,][]{gould2016}. Indeed, all free-floating planet candidates to date can be checked for putative hosts at the first AO light on any of these telescopes. While distinguishing between free-floating and wide-orbit planets in case of individual microlensing events is nearly impossible at present, the relative frequency of these objects can in principle be constrained in a statistical sense once a large sample of short-timescale events is found and characterized. 

After our earlier discoveries of ultra-short-timescale microlensing events exhibiting finite source effects \citep{mroz2018,mroz2019}, we have continued the search for similar events in data from the 2019 observing season. Here, we report the discovery of another free-floating planet candidate discovered in the microlensing event OGLE-2019-BLG-0551 ($\tE=0.381 \pm 0.017$\,day, $\thetaE=4.35 \pm 0.34$\,$\mu$as). 

\section{Data}

The microlensing event OGLE-2019-BLG-0551 occurred on a bright star located toward the Galactic bulge fields. The equatorial coordinates of the source are (R.A., Decl.)$_{\rm J2000}$ = (\ra{17}{59}{28}{74}, \dec{-28}{50}{25}{8}), which correspond to the Galactic coordinates $(l,b)=(1.626^{\circ},-2.563^{\circ})$. The source is a bright giant with a baseline magnitude of $I=13.71 \pm 0.01$ and color $V-I=2.45 \pm 0.02$.

The magnification of the source flux induced by lensing was first found by the OGLE Early Warning System \citep{udalski2003} on 2019 April 27 ($\mathrm{HJD'}=\mathrm{HJD}-2450000\approx 8600$) and the discovery was notified to the microlensing community. Two days later ($\mathrm{HJD'}\approx 8602)$, the event was independently identified by the Korea Microlensing Telescope Network \citep[KMTNet;][]{kim2016} and it was designated as KMT-2019-BLG-0519 in the KMTNet event list. The OGLE survey is conducted utilizing the 1.3\,m Warsaw telescope located at Las Campanas Observatory in Chile. The telescope is equipped with a mosaic camera that consists of 32 $2k\times4k$ detectors, yielding a 1.4\,deg$^2$ field of view with a single exposure \citep{udalski2015}. The KMTNet survey uses three identical 1.6\,m telescopes that are globally distributed at the Siding Spring Observatory in Australia (KMTA), Cerro Tololo Interamerican Observatory in Chile (KMTC), and the South African Astronomical Observatory in South Africa (KMTS). Each of the KMTNet telescopes is equipped with a camera consisting of four $9k \times 9k$ chips, yielding 4\,deg$^2$ field of view. For both surveys, images are mainly taken in the $I$ band and a small subset of images is obtained in the $V$ band for the source color measurements.

Photometry of the data was conducted using the pipelines developed by the individual survey groups: \citet{udalski2003} for the OGLE survey and \citet{albrow2009} for the KMTNet survey. These pipelines are based on the difference imaging method \citep{tomaney1996,alard1998,wozniak2000}. For the source color measurement, additional photometry was conducted using the pyDIA code \citep{albrow2017} for a subset of the KMTC data set. For the data used in the analysis, error bars from the photometry pipelines were readjusted following the routines described in \citet{yee2012} and \citet{skowron2016}.

\begin{figure}
\centering
\includegraphics[width=.9\textwidth]{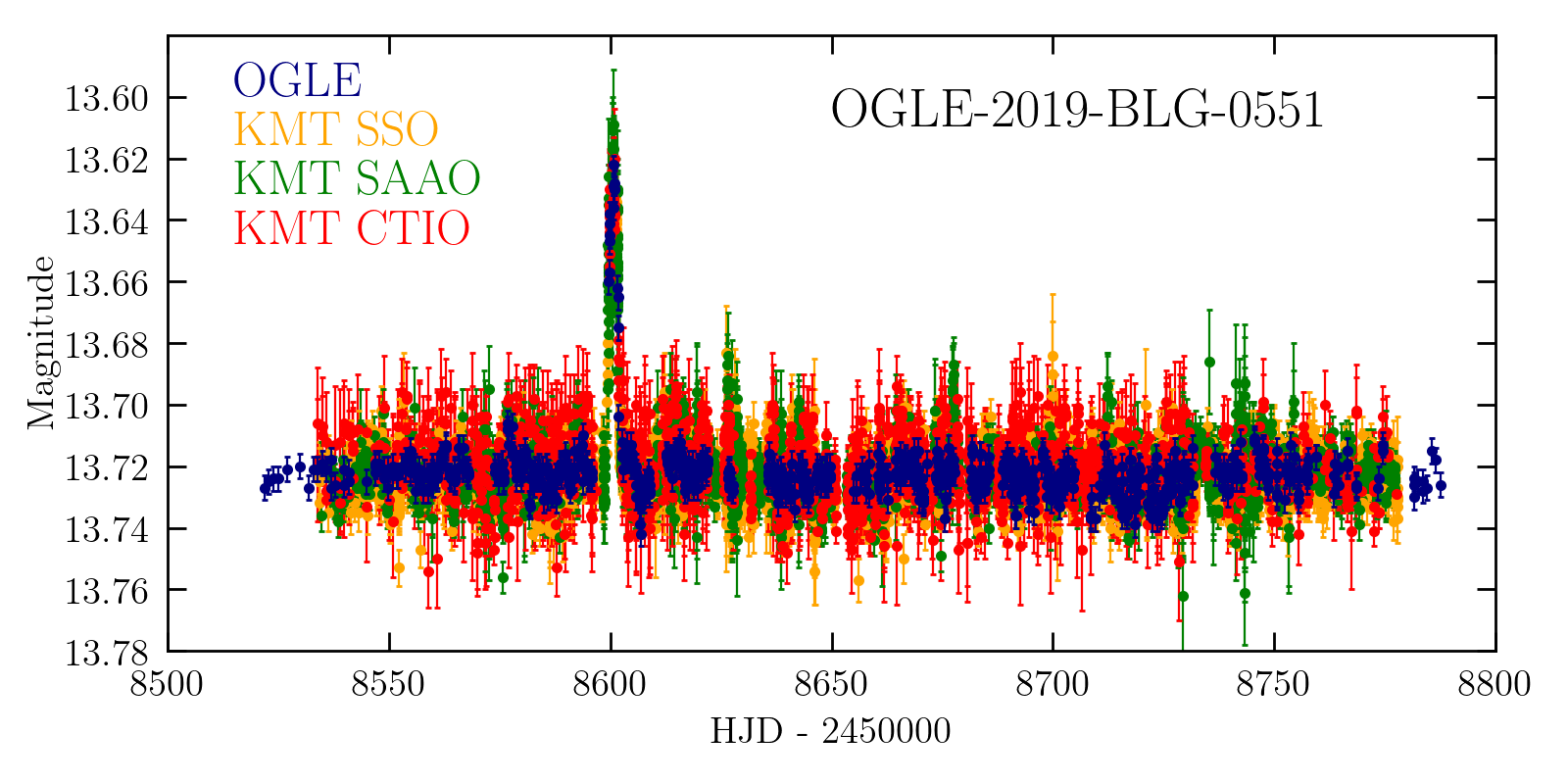}
\includegraphics[width=.9\textwidth]{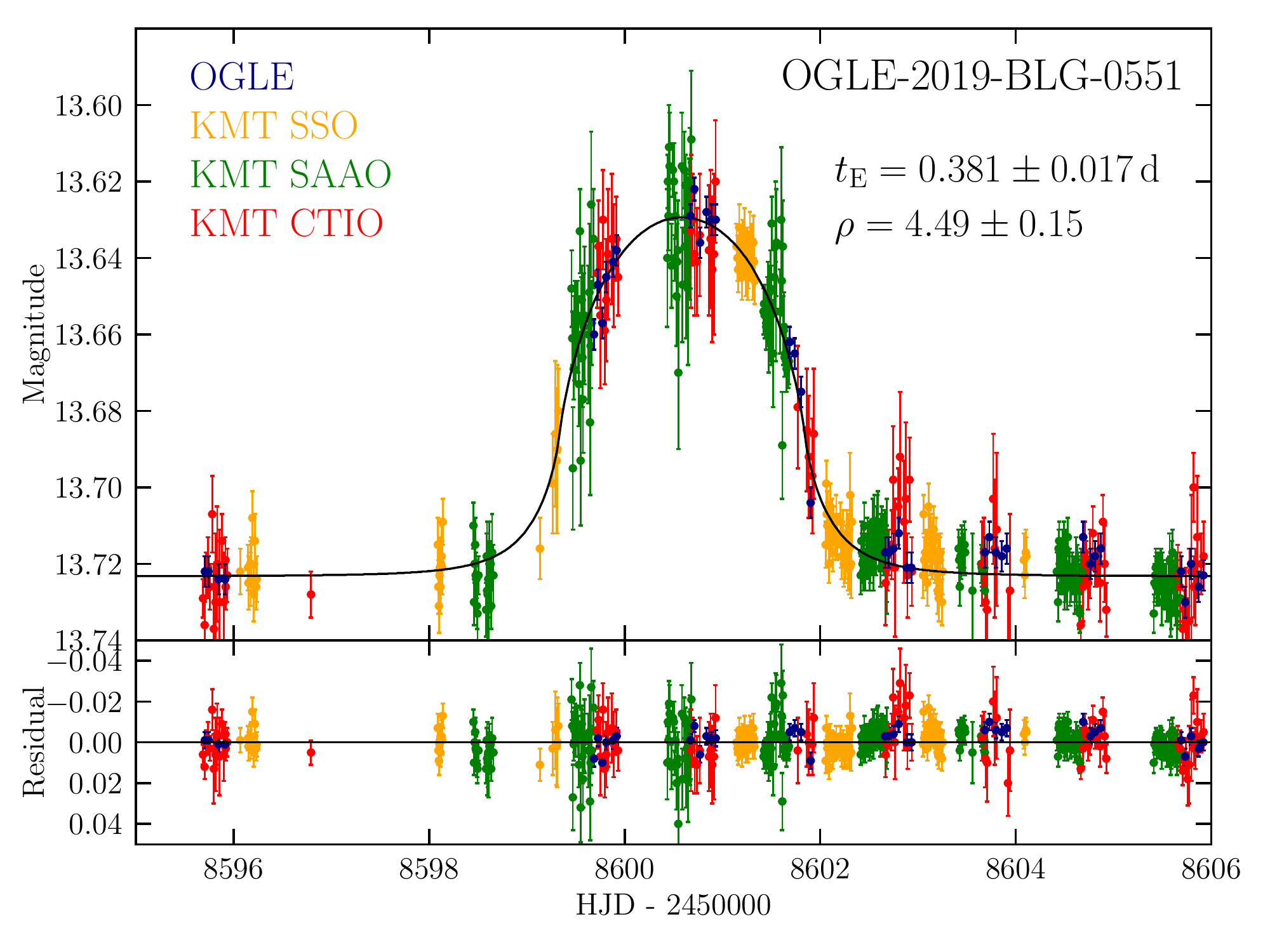}
\caption{Light curve of the microlensing event OGLE-2019-BLG-0551. Upper panel: full light curve from the 2019 observing season. Lower panel: close-up of the event. Black line is the best-fitting extended source point-lens model.}
\label{fig:single}
\end{figure}

\section{Single-lens Models}
\label{sec:espl}

The principal microlensing parameters of the event can be estimated from its light curve without the need of detailed modeling (Figure~\ref{fig:single}). The amplitude of the event (0.1\,mag) corresponds to the normalized source radius $\rho=\theta_*/\thetaE\approx4.6$ according to equation (\ref{eq:amax}), whereas its duration $\Delta t = 2.8$\,days is related to the Einstein timescale $\tE \approx \Delta t /2\rho \approx 0.3$\,day.

We modeled the light curve of the event using an extended source single-lens model with magnifications calculated using the approach of \citet{bozza2018}. In addition to $\tE$ and $\rho$, this model has two geometric parameters, $t_0$ and $u_0$, which describe the moment and separation (in Einstein radius units) during the closest approach between the lens and the center of the source. To describe the surface brightness profile of the source star, we assumed a one-parameter limb-darkening law with $\Gamma=0.56$ (Section~\ref{sec:source}). We also tested two-parameter limb-darkening profiles, but using these did not improve the quality of the fits. For the modeling we use the Markov Chain Monte Carlo (MCMC) sampler by \citet{foreman2013}.

There are additional parameters, two for each observatory, which describe the source ($\Fs$) and unmagnified blended ($\Fb$) flux. When we allowed both $\Fs$ and $\Fb$ to vary, we found that the source radius $\rho$, event timescale $\tE$, impact parameter $u_0$, and blending parameter $f_{\rm s}=\Fs/(\Fs+\Fb)$ are severely correlated. This is not surprising because the blended light would suppress the real magnification, and so---via equation (\ref{eq:amax})---would affect the normalized source radius. While low values of the blending parameter ($f_{\rm s} < 0.2$) are excluded by the data, its exact value is poorly constrained, $f_{\rm s}=0.34^{+0.44}_{-0.10}$ (see Table~\ref{tab:params}). In particular, solutions with no blending ($f_{\rm s}=1)$ are disfavored by only $\Delta\chi^2=1.2$ relative to the best-fit solution, which can be easily due to the noise in the data.

In the best-fit solution, the source and blend have $I$-band magnitudes of 15.25 and 14.03, respectively. Such bright stars are relatively rare and the prior probability of having them blended is extremely small. For example, image-level simulations of \citet{mroz2019b} (their Figure~7) showed that the distribution of $f_{\rm s}$ of bright stars is bimodal: either the entire flux comes from the source ($f_{\rm s}\approx 1$) or the source is significantly fainter than the blend ($f_{\rm s}\approx 0$) (see also \citealt{smith2007} and \citealt{wyrzykowski2015}).

We combined OGLE and \textit{Hubble Space Telescope} (\textit{HST}) observations \citep{brown2009,brown2010} of a nearby \textit{HST} ``Stanek'' field to derive the empirical distribution of blending parameter of bright stars. We cross-matched individual stars detected on the \textit{HST} image with stars on the OGLE reference image. We subsequently calculated the ratio of their flux $F_{\rm HST}$ to the total flux of the object detected on OGLE template image $F_{\rm OGLE}$. The blending parameter is simply the ratio $f_{\rm s}=F_{\rm HST}/F_{\rm OGLE}$. We used this empirical distribution as a prior on $f_{\rm s}= 1.00 \pm 0.06$. The resulting parameters (Table~\ref{tab:params}) are much better constrained and are consistent with our approximate estimates based on the amplitude and duration of the event. 

We note that the physical interpretation of both models (with and without the prior on blending) is nearly identical ($\tE \approx 0.4-0.5$\,day and $\thetaE\approx 4.2-4.4$\,$\mu$as). In particular, it is noteworthy that the angular Einstein radii are virtually the same in both models. We explain these facts in Section~\ref{sec:thetaE}.

\begin{table}
\caption{Best-fitting Single-lens Model Parameters}
\centering
\begin{tabular}{lrr}
\hline \hline
& Single Lens & Single Lens \\
& w/o Blend Prior & w/ Blend Prior \\
\hline
\multicolumn{3}{l}{Microlensing model} \\
\hline
$t_0$ (HJD') & $8600.584 \pm 0.011$      & $8600.586 \pm 0.011$ \\
$\tE$ (days) & $0.505^{+0.071}_{-0.102}$ & $0.381 \pm 0.017$    \\
$u_0$        & $1.01^{+1.49}_{-0.71}$     & $3.02 \pm 0.16$      \\
$\rho$       & $2.73^{+1.27}_{-0.45}$    & $4.49 \pm 0.15$      \\
$I_{\rm s}$  & $14.91^{+0.37}_{-0.90}$   & $13.73 \pm 0.07$     \\
$f_{\rm s}$  & $0.34^{+0.44}_{-0.10}$    & $0.99 \pm 0.06$      \\
$\chi^2$/dof & $11889/11006$ & \dots \\
\hline
\multicolumn{3}{l}{Source star} \\
\hline
$I_{\rm s,0}$             & $13.79^{+0.37}_{-0.90}$ & $12.61 \pm 0.06$\\ 
$(V-I)_{\rm s,0}$         & $1.49 \pm 0.02$ & $1.49 \pm 0.02$\\
$T_{\rm eff}$ (K)         & $4000 \pm 200$ & $4000 \pm 200$ \\
$\Gamma$ (limb-darkening, $I$ band) & 0.56 & 0.56 \\
$\theta_*$ ($\mu$as)      & $11.4 ^{+5.9}_{-1.8}$  & $19.5 \pm 1.6$    \\
\hline
\multicolumn{3}{l}{Physical parameters} \\
\hline
$\thetaE$ ($\mu$as)       & $4.23 \pm 0.34$ & $4.35 \pm 0.34$   \\
$\murel$ (mas\,yr$^{-1}$) & $3.01^{+0.86}_{-0.36}$   & $4.17 \pm 0.35$     \\
\hline
\end{tabular}
\label{tab:params}
\end{table}

\begin{figure}
\centering
\includegraphics[width=\textwidth]{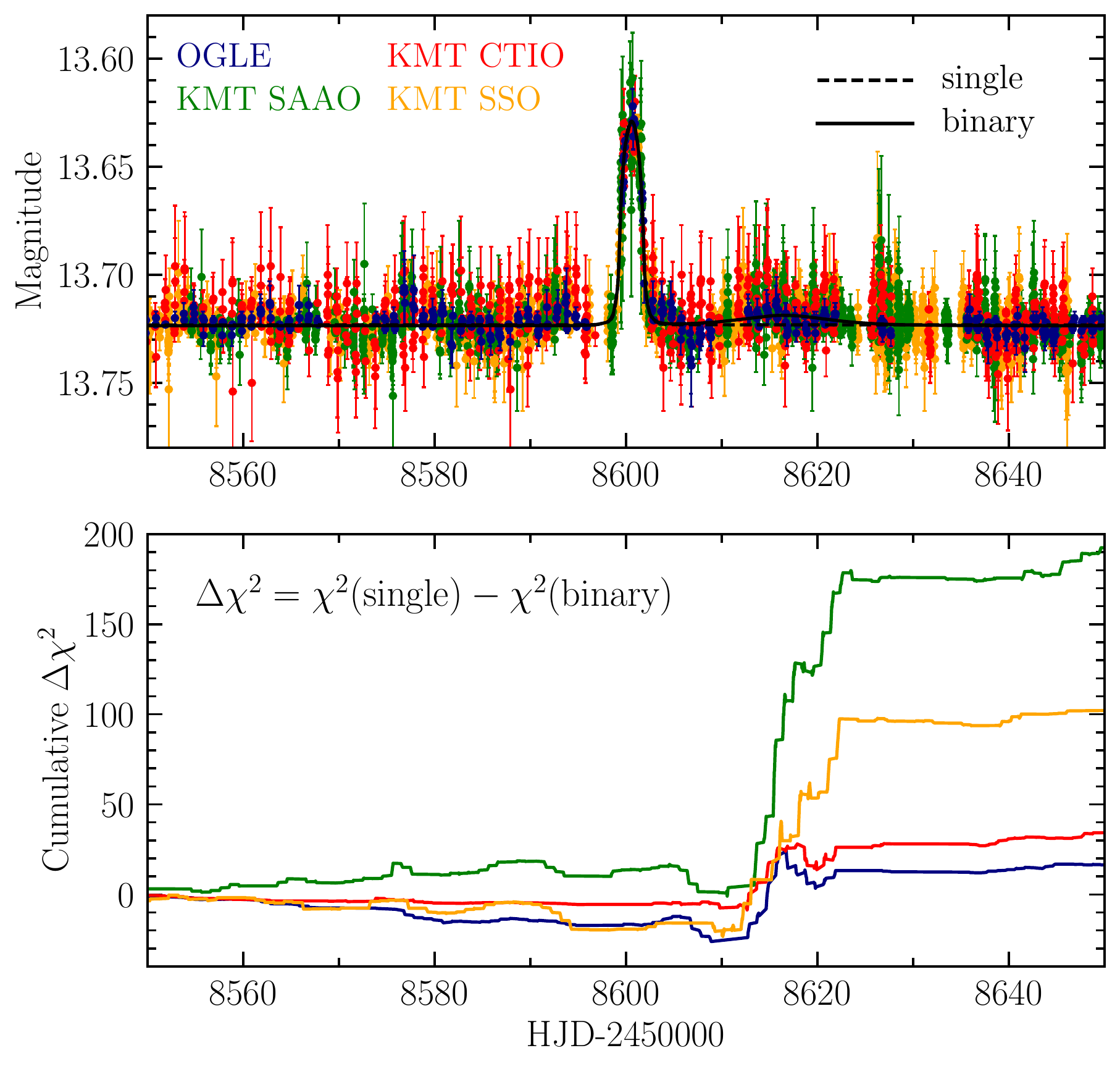}
\caption{Upper panel: comparison between the single- (dashed line) and binary-lens (solid line) models. Lower panel: cumulative distribution of $\Delta\chi^2$ between these models.}
\label{fig:binary}
\end{figure}

\begin{figure}
\centering
\includegraphics[width=.5\textwidth]{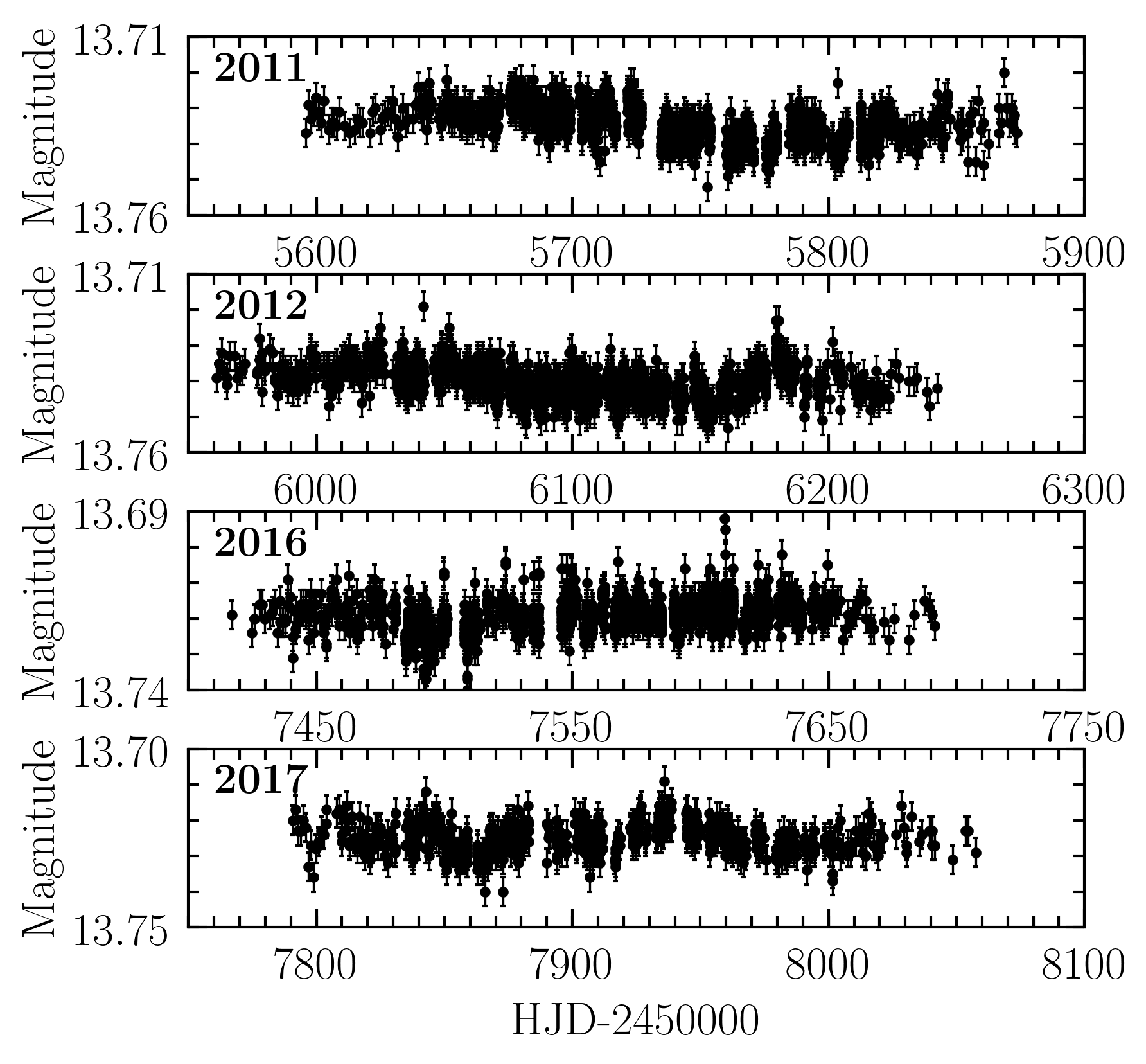}
\caption{Archival OGLE light curve of OGLE-2019-BLG-0551 from observing seasons 2011, 2012, 2016, and 2017. The source star exhibits low-amplitude ($<0.01$\,mag) variability on timescales of a few tens of days.}
\label{fig:archival}
\end{figure}

\section{Binary-lens Models and Variability of the Source}

We also carried out an extensive search for binary-lens solutions. The event does not show very clear departures from the point-lens model, but it is important to search for a possible host of the planet, i.e., fit the binary-lens model. To parameterize the binary-lens model we need three additional parameters: a mass ratio ($q$) and two parameters defining the geometry of the event -- projected separation in units of $\theta_{\rm E}$ ($s$) and angle between the lens axis and the source trajectory ($\alpha$). For the binary-lens model fitting, we use the parameters that are directly constrained by the data. Thus, we define parameters $t_0$ and $u_0$ relative to the approximate position of the planetary caustic \citep[$s-1/s$ relative to the host star;][]{han2006}. We also use $t_{\star}=\rho\tE$ (source radius crossing time) instead of $\rho$, and $t_{{\rm E, planet}} = \tE \sqrt{q/(1+q)}$ (Einstein timescale corresponding to the mass of planet) instead of the Einstein timescale relative to the total mass of the binary lens ($\tE$). 

We calculate finite-source binary-lens magnifications using the method by \citet{bozza2010} and \citet{bozza2018} as implemented in MulensModel package \citep{poleski_yee2019}. We define a wide grid in $(s, q)$ and run separate MCMC \citep{foreman2013} chains with fixed $(s, q)$ at every point in the grid. We ignore limb-darkening effects in the grid calculations. The local solutions from the grid search were once more refined using MCMC but this time all parameters were fitted and limb-darkening was taken into account.

There are two degenerate binary-lens models differing by $\Delta\chi^2=8.5$ because the source can pass the planetary caustic on either of the two sides \citep{gaudi1997}. Here we report results for the model with smaller $\chi^2$. The best-fitting binary-lens model is nearly identical to the single-lens model, except of a low-amplitude ``bump'' ($\sim 0.005$\,mag) due to the host star with a maximum $\sim 15$\,days after the main event. See the upper panel of Figure~\ref{fig:binary}. Formally, the binary-lens model is preferred over the single-lens model by $\Delta\chi^2=381$ (Table~\ref{tab:params_binary}). However, the source exhibits irregular low-amplitude variability in the archival data (Figure~\ref{fig:archival}), and such variations are typical for red giant stars. The amplitude and timescales of these variations are similar to those of the ``bump'' after the event. This raises the possibility that much (if not all) of the $\chi^2$ improvement is due to variability of the source.

To check this possibility, we plotted the cumulative distribution of $\Delta\chi^2$ between the best-fitting binary- and single-lens models for the individual data sets (lower panel of Figure~\ref{fig:binary}), which contribute to the $\chi^2$ improvement by $\Delta\chi^2=31$ (OGLE), $\Delta\chi^2=34$ (KMTC), $\Delta\chi^2=183$ (KMTS), and $\Delta\chi^2=133$ (KMTA). This indicates that the ``bump'' in the light curve is indeed real. The largest $\chi^2$ improvement can be attributed to data points collected during $8612<\mathrm{HJD}'<8622$. However, according to the binary model, measurements taken during $8605<\mathrm{HJD}'<8612$ should also be slightly magnified, which contradicts the OGLE data (and hence the cumulative $\Delta\chi^2$ decreases during that time). The overall shape of the cumulative $\Delta\chi^2$ distribution suggests that the shape of the ``bump'' in the light curve of the event does not match that expected from the microlensing model and is likely due to low-level variability of the source star. 

To check how well the source variability may mimic microlensing signal from the host star, we fitted simple point-lens point-source models to the archival OGLE light curves from seasons 2011 to 2018 (Figure~\ref{fig:archival}). We found that they may be formally preferred over the constant brightness models by as large as $\Delta\chi^2=588$ (``bump'' at $\mathrm{HJD}' \approx 6024$), 193 ($\mathrm{HJD}' \approx 6180$), or 127 ($\mathrm{HJD}' \approx 7938$). This demonstrates that the $\chi^2$ improvement due to the ``bump'' at $\mathrm{HJD}' \approx 8615$ (which is a sum of contributions from OGLE and three KMTNet observatories) can be easily explained by the variability of the source. 

We also note that the parameters of the binary lens-model ($\tE= 2.17 \pm 0.32$\,days, $q= 0.043 \pm 0.011$) correspond to a priori unlikely physical configuration consisting of a super Jupiter-mass planet (or a brown dwarf) orbited by a Neptune-mass object. This supports the idea that the ``bump'' in the OGLE-2019-BLG-0551 light curve is caused by the variability of the source star.

\begin{table}
\caption{Best-fitting Binary-lens Model Parameters}
\centering
\begin{tabular}{lr}
\hline \hline
Parameter & Value \\
\hline
$t_0$ (HJD') & $8615.75 \pm 0.51$ \\
$\tE$ (days) & $2.17 \pm 0.32$ \\
$u_0$        & $3.41 \pm 0.47$ \\
$t_*$ (days) & $0.67 \pm 0.15$ \\
$q$          & $0.043 \pm 0.011$ \\
$s$          & $8.39 \pm 1.06$ \\
$\alpha$     & $3.631 \pm 0.040$ \\
\hline
$\chi^2$/dof & $11508/11003$ \\
\hline
\end{tabular}
\label{tab:params_binary}

Note: All parameters are relative to the center of mass of the lens.
\end{table}

\begin{figure}
\centering
\includegraphics[width=.8\textwidth]{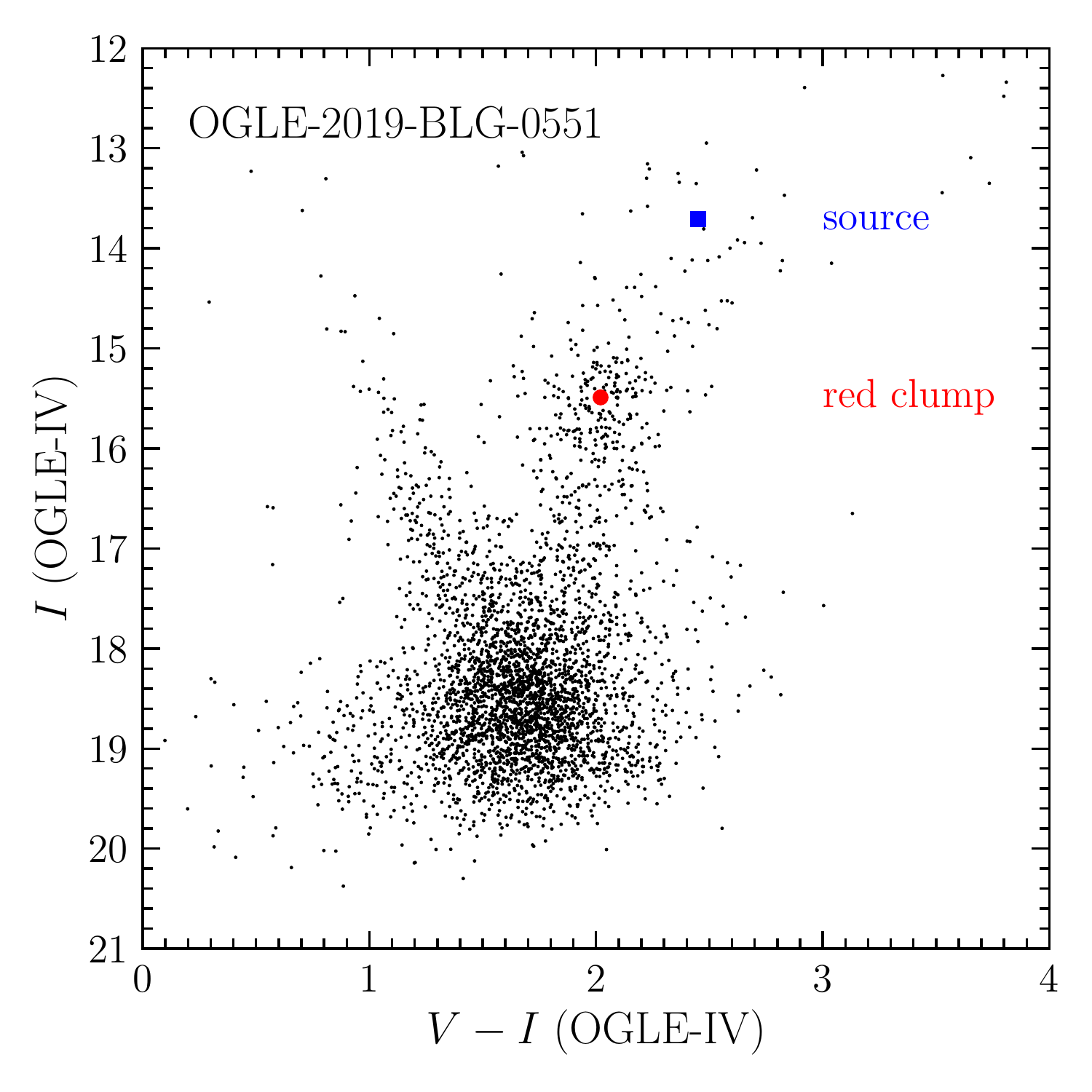}
\caption{Calibrated color--magnitude diagram of stars within $2'$ of OGLE-2019-BLG-0551.}
\label{fig:cmd}
\end{figure}

\section{Physical parameters}
\label{sec:source}

\subsection{Source star}
\label{sec:cmd}

With the normalized source radius $\rho$ measured from the light curve of the event, we estimate the angular Einstein radius using the relation 
\begin{equation}
\thetaE = \frac{\theta_*}{\rho}.
\end{equation}
For this, we first estimate the angular source radius, $\theta_*$, based on the de-reddened source color $(V-I)_{\rm s,0}$ and brightness $I_{\rm s,0}$, using the standard method of \citet{yoo2004}. We first locate the source in the calibrated color--magnitude diagram (CMD), measure the offsets of the source in color, $\Delta(V-I)$, and brightness, $\Delta I$, from the centroid of the red giant clump in the CMD, and then estimate $(V-I)_{\rm s,0}$ and $I_{\rm s,0}$ using the relation 
\begin{equation}
(V-I,I)_{\rm s,0} = (V-I,I)_{\rm RC,0} + \Delta(V-I,I).
\end{equation}
Here $(V-I,I)_{\rm RC,0}=(1.06,14.37)$ denote the known values of the de-reddened color and brightness of the red clump centroid \citep{bensby2011,nataf2013}.

In Figure~\ref{fig:cmd}, we present the locations of the source (blue square) and the red clump centroid (red circle) on the calibrated CMD of stars around the source. The color and brightness of the source are $(V-I,I)=(2.45 \pm 0.02,13.73\pm0.06)$ and those of the red clump centroid are $(V-I,I)_{\rm RC}=(2.02,15.49)$. Here, we assumed that the source color is the same as the color of the baseline object. To test this assumption, we used pyDIA reductions of a subset of KMTC data covering the event both in the $I$ and $V$ bands to calculate the source color for each link of our MCMC chain. The mean instrumental source color in our model is $V-I = 2.56 \pm 0.03$, whereas the mean instrumental color of the ``baseline object'' is $V-I = 2.57 \pm 0.02$, which justifies our assumption. The agreement of baseline object and source colors argues extremely strongly against significant blended light, since the blend would have to have essentially the same color as the source, which is extraordinarily red.

With the measured offsets in color and brightness, $\Delta(V-I,I) = (0.43,-1.76)$, we can estimate that the source has a de-reddened color and brightness of
\begin{equation}
(V-I,I)_{\rm s,0} = (1.49 \pm 0.02,12.61\pm 0.06).
\end{equation}
The measured color and brightness indicate that the source is a giant star with a spectral type K4 and effective temperature $T_{\rm eff} = 4000 \pm 200$\,K \citep{houdashelt2000}. The corresponding limb-darkening coefficients $\Gamma$ are 0.56 and 0.81, in the $I$ and $V$ bands, respectively \citep{claret2011}.
With the measured $(V-I)_{\rm s,0}$ and $I_{\rm s,0}$, we estimate the angular source radius from the color--surface brightness relation of \citet{adams2018} for giants: $\theta_* = 19.55 \pm 1.57 \,\mu\mathrm{as}$ (which is valid in the range of $-0.01<V-I<1.74$).

\subsection{Angular Einstein radius}
\label{sec:thetaE}

The angular Einstein radius is estimated as
\begin{equation}
\thetaE = \frac{\theta_*}{\rho}= 4.35 \pm 0.34\,\mu\mathrm{as},
\end{equation}
which makes it the second lowest known $\thetaE$ of short-timescale events (after OGLE-2012-BLG-1323: $\thetaE = 2.37 \pm 0.10$\,$\mu$as).
Together with the measured event timescale, the relative lens-source proper motion is estimated as 
\begin{equation}
\murel = \frac{\thetaE}{\tE}= 4.18 \pm 0.35 \,\mathrm{mas\,yr}^{-1}.
\end{equation}
All physical parameters of the lens are summarized in Table~\ref{tab:params}.

We noted in Section~\ref{sec:espl} that the values of the angular Einstein radius in our single-lens models with and without prior on blending are virtually the same. Here we explore mathematical reasons explaining this coincidence. Let $I_0$ and $I_{\rm s}$ be the baseline and source magnitudes, respectively. From the Pogson's law, $I_0-I_{\rm s}=2.5\log f_{\rm s}$ \citep{pogson}, where $f_{\rm s}$ is the dimensionless blending parameter (Section~\ref{sec:espl}). The only way the brightness of the source affects the physical parameters of the model is via angular size of the source,
\begin{equation}
\theta_* = \theta_{*,0}\cdot 10^{0.2(I_0-I_{\rm s})} = \theta_{*,0}\sqrt{f_{\rm s}},
\label{eq:eq1}
\end{equation}
where $\theta_{*,0}$ is the angular radius of the source corresponding to no blending. (We assume that blending does not affect the color of the source. This will always be the case when the source color is determined from regression. In the present case, we are using the color of the baseline object, but we found in Section~\ref{sec:cmd} that it is consistent with the source color measured from the microlensing model.) 

Blending tends to lower the amplitude of the event. Equation~(\ref{eq:amax}), in the presence of blending and assuming no limb-darkening, can be rewritten as
\begin{equation}
A_{\rm max} = 1+2f_{\rm s}\left(\frac{\thetaE}{\theta_*}\right)^2 = 1+\frac{2 f_{\rm s}}{\rho^2}.
\end{equation}
The amplitude of the event is well measured from the light curve. Therefore
\begin{equation}
\rho = \sqrt{\frac{2 f_{\rm s}}{A_{\rm max}-1}}=\rho_0 \sqrt{f_{\rm s}},
\label{eq:eq2}
\end{equation}
where $\rho_0$ denotes the normalized source radius corresponding to no blending. Equations~(\ref{eq:eq1}) and~(\ref{eq:eq2}) explain why the value of the angular Einstein radius $\thetaE = \theta_*/\rho = \theta_{*,0}/\rho_0$ is insensitive to the changes of the blending parameter.

In fact, the independence of the $\thetaE$ estimate from blending stems from a much simpler principle. If a lens with $\thetaE$ transits a source with $\theta_* \gg \thetaE$, then to zeroth order, the excess flux $\Delta F$ is given by
\begin{equation}
\Delta F = 2\pi S_0 \thetaE^2,
\end{equation}
where $S_0$ is the surface brightness of the source at the lens center. Therefore, if $S_0$ is known (from the color of the source), then
\begin{equation}
\thetaE = \sqrt{\frac{S_0}{2\pi\Delta F}}
\end{equation}
can be derived from purely empirical quantities, without any knowledge of either $\theta_*$ or $\rho$ (provided it is known that $\theta_* \gg \thetaE$).

\begin{figure}
\centering
\includegraphics[width=.7\textwidth]{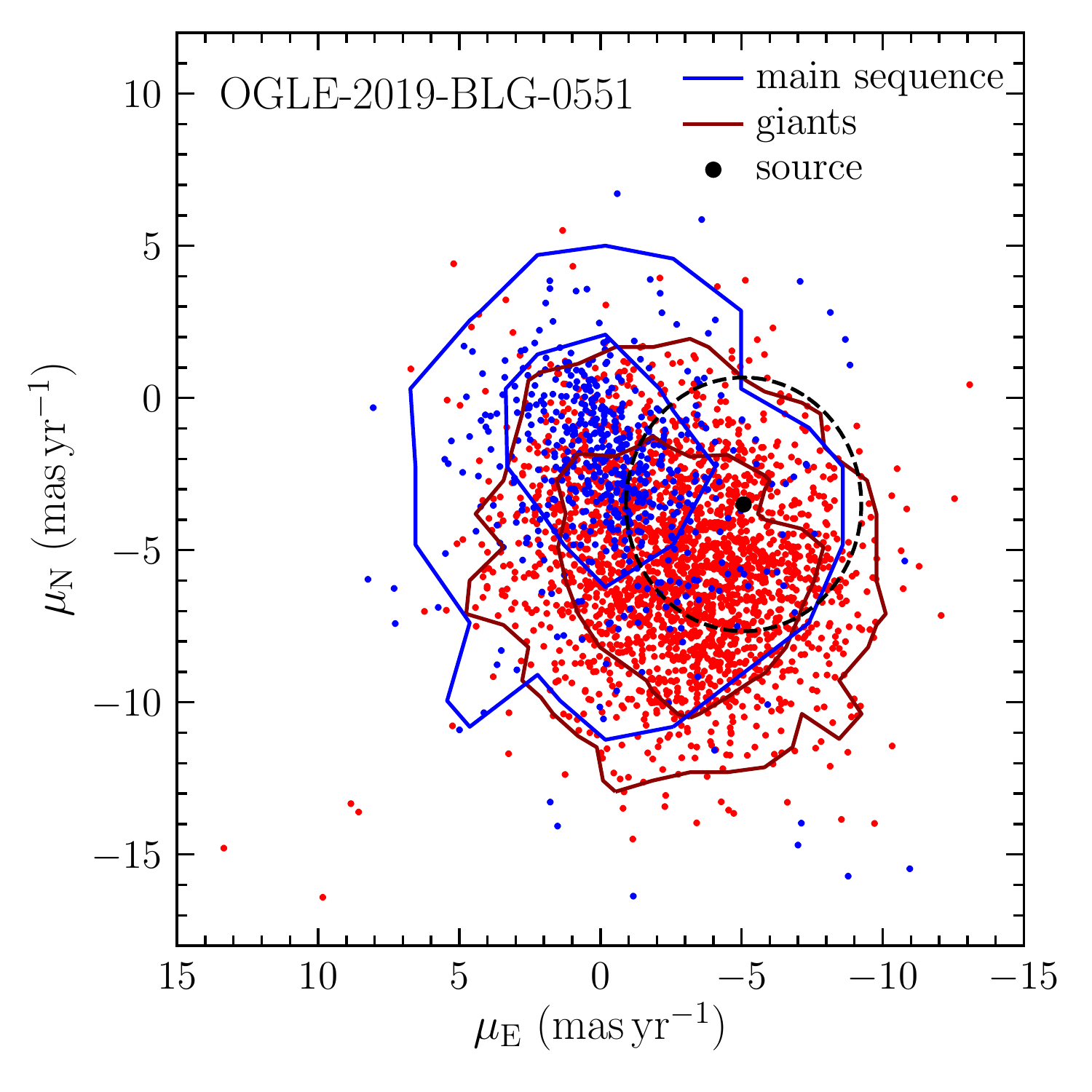}
\caption{\textit{Gaia} DR2 proper motions of stars within $4'$ of OGLE-2019-BLG-0551. Blue contours correspond to main-sequence stars (which represent the Galactic disk population) and red contours to giants (bulge population). The source is marked with a black dot. The lens should be located on the dashed circle, which radius corresponds to the relative lens-source proper motion of $4.17 \pm 0.35$\,mas\,yr$^{-1}$. Solid contours enclose 68\% and 95\% of all objects.}
\label{fig:pm}
\end{figure}

\subsection{Proper motion of the source}

Additional information about the source and lens can be obtained from the second \textit{Gaia} data release \citep[DR2;][]{gaia2016,gaia2018}. Figure~\ref{fig:pm} presents \textit{Gaia} DR2 proper motions of stars within $4'$ of OGLE-2019-BLG-0551. Blue dots and contours correspond to main-sequence stars (which represent the Galactic disk population), whereas giants (Galactic bulge population) are marked in red. The proper motion of the source ($\mu_{\rm E} = -5.07 \pm 0.20$\,mas\,yr$^{-1}$, $\mu_{\rm N} = -3.50 \pm 0.15$\,mas\,yr$^{-1}$) is consistent with that of bulge stars. 

Thanks to the detection of finite source effects, we were able to measure the relative lens-source proper motion of $4.17 \pm 0.35$\,mas\,yr$^{-1}$. Thus, the lens should be located on the dashed circle in Figure~\ref{fig:pm}, and the proper motion of the lens is consistent with that of both Galactic disk and bulge stars.

\section{Discussion and conclusions}

After the discovery of a handful of ultra-short-timescale microlensing events exhibiting strong finite source effects in the archival data \citep{mroz2018,mroz2019}, we continue the search for similar events among microlenses detected in real time by the OGLE Early Warning System \citep{udalski2003}. OGLE-2019-BLG-0551 was identified as a promising candidate soon after its public announcement in 2019 April, based on its short duration and low amplitude. However, the detailed analysis of the event was postponed until the end of the observing season to track the evolution of the light curve, as possible signatures from the putative host star may become apparent weeks to months after the main event. 

The light curve of the event can be well described by the extended source single-lens model with an Einstein timescale $\tE=0.381 \pm 0.017$\,day that is much shorter, and an angular Einstein radius $\thetaE=4.35 \pm 0.34$\,$\mu$as that is much smaller than those of typical microlensing events produced by stellar-mass lenses ($\tE \sim 20$\,days, $\thetaE \sim 0.3$\,mas). In fact, this lens has the second lowest $\thetaE$ of known short-timescale events. This indicates that the lens is likely to be a planetary-mass object, although its mass cannot be unambiguously determined because the relative lens-source parallax $\pirel$ is unknown:
\begin{equation}
M = \frac{\thetaE^2}{\kappa \pirel}=7.7\,M_{\oplus} \frac{0.1\,\mathrm{mas}}{\pirel}.
\end{equation}
The lens may be a sub-Neptune-mass planet in the Galactic disk ($\pirel\approx 0.1\,$mas) or a Saturn-mass object located in the Galactic bulge ($\pirel\approx 0.01\,$mas). When $\tE$ and $\thetaE$ are known from the light curve, the mass of the lens can be in principle constrained using the Bayesian analysis based on priors on the mass function of lenses and priors on the Galactic structure and kinematics. Because of the extreme nature of this event, we chose not to carry out the Bayesian mass estimate.

We cannot rule out the possibility that the lens orbits a distant companion star. We conducted an extensive search for possible binary-lens solutions, which could in principle have revealed the signal from the companion star. Although we found that the binary-lens model is formally preferred over the single-lens model by $\Delta\chi^2=381$, we argue that the entire $\chi^2$ improvement can be explained by the low-level variability of the source, which is apparent in the archival OGLE light curve. Thus, we do not find any significant evidence for the host star.

If there exists a host star and it is luminous, it may be detected in the future, when the lens and source separate. Because the source star is extremely bright, a separation of about 2 FWHM is required. Such separation will be achieved after
\begin{equation}
\delta t = 3.4\,\mathrm{yr}\left(\frac{\murel}{4.18\,\mathrm{mas\,yr}^{-1}}\right)^{-1} \left(\frac{\lambda}{1.1\,\mu\mathrm{m}}\right)\left(\frac{D}{39\,\mathrm{m}}\right)^{-1},
\end{equation}
where $\lambda$ is the wavelength of the observation and $D$ is the diameter of the mirror. Thus, this method can be applied at the first light of AO from any of the planned extremely large telescopes.

\section*{Acknowledgements}

P.M. acknowledges support from the National Science Center, Poland (grant ETIUDA 2018/28/T/ST9/00096). R.P. was supported by Polish National Agency for Academic Exchange via Polish Returns 2019 grant. The OGLE project has received funding from the National Science Centre, Poland, grant MAESTRO 2014/14/A/ST9/00121 to A.U. Work by A.G. was supported by AST-1516842 from the US NSF and by JPL grant 1500811. Work by C.H. was supported by the grants of National Research Foundation of Korea (2017R1A4A1015178 and 2019R1A2C2085965).
This research has made use of the KMTNet system operated by the Korea Astronomy and Space Science Institute (KASI) and the data were obtained at three host sites of CTIO in Chile, SAAO in South Africa, and SSO in Australia.

\bibliographystyle{aasjournal}
\bibliography{sample}

\begin{thebibliography}{}
\expandafter\ifx\csname natexlab\endcsname\relax\def\natexlab#1{#1}\fi

\bibitem[{{Adams} {et~al.}(2018){Adams}, {Boyajian}, \& {von
  Braun}}]{adams2018}
{Adams}, A.~D., {Boyajian}, T.~S., \& {von Braun}, K. 2018, \mnras, 473, 3608

\bibitem[{{Alard} \& {Lupton}(1998)}]{alard1998}
{Alard}, C., \& {Lupton}, R.~H. 1998, \apj, 503, 325

\bibitem[{{Albrow}(2017)}]{albrow2017}
{Albrow}, M. 2017, MichaelDAlbrow/pyDIA: Initial Release on Github., , ,
  doi:10.5281/zenodo.268049

\bibitem[{{Albrow} {et~al.}(2009){Albrow}, {Horne}, {Bramich}, {Fouqu{\'e}},
  {Miller}, {Beaulieu}, {Coutures}, {Menzies}, {Williams}, {Batista},
  {Bennett}, {Brillant}, {Cassan}, {Dieters}, {Dominis Prester}, {Donatowicz},
  {Greenhill}, {Kains}, {Kane}, {Kubas}, {Marquette}, {Pollard}, {Sahu},
  {Tsapras}, {Wambsganss}, \& {Zub}}]{albrow2009}
{Albrow}, M.~D., {Horne}, K., {Bramich}, D.~M., {et~al.} 2009, \mnras, 397,
  2099

\bibitem[{{Andrews} {et~al.}(2018){Andrews}, {Huang}, {P{\'e}rez}, {Isella},
  {Dullemond}, {Kurtovic}, {Guzm{\'a}n}, {Carpenter}, {Wilner}, {Zhang}, {Zhu},
  {Birnstiel}, {Bai}, {Benisty}, {Hughes}, {{\"O}berg}, \&
  {Ricci}}]{andrews2018}
{Andrews}, S.~M., {Huang}, J., {P{\'e}rez}, L.~M., {et~al.} 2018, \apjl, 869,
  L41

\bibitem[{{Baron} {et~al.}(2019){Baron}, {Lafreni{\`e}re}, {Artigau},
  {Gagn{\'e}}, {Rameau}, {Delorme}, \& {Naud}}]{baron2019}
{Baron}, F., {Lafreni{\`e}re}, D., {Artigau}, {\'E}., {et~al.} 2019, \aj, 158,
  187

\bibitem[{{Bennett} \& {Rhie}(1996)}]{bennett1996}
{Bennett}, D.~P., \& {Rhie}, S.~H. 1996, \apj, 472, 660

\bibitem[{{Bensby} {et~al.}(2011){Bensby}, {Ad{\'e}n}, {Mel{\'e}ndez}, {Gould},
  {Feltzing}, {Asplund}, {Johnson}, {Lucatello}, {Yee}, {Ram{\'{\i}}rez},
  {Cohen}, {Thompson}, {Bond}, {Gal-Yam}, {Han}, {Sumi}, {Suzuki}, {Wada},
  {Miyake}, {Furusawa}, {Ohmori}, {Saito}, {Tristram}, \&
  {Bennett}}]{bensby2011}
{Bensby}, T., {Ad{\'e}n}, D., {Mel{\'e}ndez}, J., {et~al.} 2011, \aap, 533,
  A134

\bibitem[{{Bozza}(2010)}]{bozza2010}
{Bozza}, V. 2010, \mnras, 408, 2188

\bibitem[{{Bozza} {et~al.}(2018){Bozza}, {Bachelet}, {Bartoli{\'c}}, {Heintz},
  {Hoag}, \& {Hundertmark}}]{bozza2018}
{Bozza}, V., {Bachelet}, E., {Bartoli{\'c}}, F., {et~al.} 2018, \mnras, 479,
  5157

\bibitem[{{Brown} {et~al.}(2009){Brown}, {Sahu}, {Zoccali}, {Renzini},
  {Ferguson}, {Anderson}, {Smith}, {Bond}, {Minniti}, {Valenti}, {Casertano},
  {Livio}, {Panagia}, {Vanden Berg}, \& {Valenti}}]{brown2009}
{Brown}, T.~M., {Sahu}, K., {Zoccali}, M., {et~al.} 2009, \aj, 137, 3172

\bibitem[{{Brown} {et~al.}(2010){Brown}, {Sahu}, {Anderson}, {Tumlinson},
  {Valenti}, {Smith}, {Jeffery}, {Renzini}, {Zoccali}, {Ferguson},
  {VandenBerg}, {Bond}, {Casertano}, {Valenti}, {Minniti}, {Livio}, \&
  {Panagia}}]{brown2010}
{Brown}, T.~M., {Sahu}, K., {Anderson}, J., {et~al.} 2010, \apjl, 725, L19

\bibitem[{{Claret} \& {Bloemen}(2011)}]{claret2011}
{Claret}, A., \& {Bloemen}, S. 2011, \aap, 529, A75

\bibitem[{{Flock} {et~al.}(2015){Flock}, {Ruge}, {Dzyurkevich}, {Henning},
  {Klahr}, \& {Wolf}}]{flock2015}
{Flock}, M., {Ruge}, J.~P., {Dzyurkevich}, N., {et~al.} 2015, \aap, 574, A68

\bibitem[{{Foreman-Mackey} {et~al.}(2013){Foreman-Mackey}, {Hogg}, {Lang}, \&
  {Goodman}}]{foreman2013}
{Foreman-Mackey}, D., {Hogg}, D.~W., {Lang}, D., \& {Goodman}, J. 2013, \pasp,
  125, 306

\bibitem[{{Gaia Collaboration} {et~al.}(2016){Gaia Collaboration}, {Prusti},
  {de Bruijne}, {Brown}, {Vallenari}, {Babusiaux}, {Bailer-Jones}, {Bastian},
  {Biermann}, {Evans}, \& et~al.}]{gaia2016}
{Gaia Collaboration}, {Prusti}, T., {de Bruijne}, J.~H.~J., {et~al.} 2016,
  \aap, 595, A1

\bibitem[{{Gaia Collaboration} {et~al.}(2018){Gaia Collaboration}, {Brown},
  {Vallenari}, {Prusti}, {de Bruijne}, {Babusiaux}, {Bailer-Jones}, {Biermann},
  {Evans}, {Eyer}, \& et~al.}]{gaia2018}
{Gaia Collaboration}, {Brown}, A.~G.~A., {Vallenari}, A., {et~al.} 2018, \aap,
  616, A1

\bibitem[{{Gaudi} \& {Gould}(1997)}]{gaudi1997}
{Gaudi}, B.~S., \& {Gould}, A. 1997, \apj, 486, 85

\bibitem[{{Gould}(1994)}]{gould1994}
{Gould}, A. 1994, \apjl, 421, L71

\bibitem[{{Gould}(2016)}]{gould2016}
{Gould}, A. 2016, \jkas, 49, 123

\bibitem[{{Gould} \& {Gaucherel}(1997)}]{gould1997}
{Gould}, A., \& {Gaucherel}, C. 1997, \apj, 477, 580

\bibitem[{{Han}(2006)}]{han2006}
{Han}, C. 2006, \apj, 638, 1080

\bibitem[{{Han} {et~al.}(2005){Han}, {Gaudi}, {An}, \& {Gould}}]{han2005}
{Han}, C., {Gaudi}, B.~S., {An}, J.~H., \& {Gould}, A. 2005, \apj, 618, 962

\bibitem[{{Han} \& {Kang}(2003)}]{han2003}
{Han}, C., \& {Kang}, Y.~W. 2003, \apj, 596, 1320

\bibitem[{{Han} {et~al.}(2020{\natexlab{a}}){Han}, {Lee}, {Udalski}, {Gould},
  {Bond}, {Bozza}, {Albrow}, {Chung}, {Hwang}, {Jung}, {Ryu}, {Shin},
  {Shvartzvald}, {Yee}, {Zang}, {Cha}, {Kim}, {Kim}, {Kim}, {Lee}, {Lee},
  {Park}, {Pogge}, {Jee}, {Kim}, {KMTnet Collaboration}, {Mr{\'o}z},
  {Szyma{\'n}ski}, {Skowron}, {Poleski}, {Soszy{\'n}ski}, {Pietrukowicz},
  {Koz{\l}owski}, {Ulaczyk}, {Rybicki}, {Iwanek}, {Wrona}, {OGLE
  Collaboration}, {Abe}, {Barry}, {Bennett}, {Bhattacharya}, {Donachie},
  {Fujii}, {Fukui}, {Itow}, {Hirao}, {Kamei}, {Kondo}, {Koshimoto}, {Li},
  {Matsubara}, {Muraki}, {Miyazaki}, {Nagakane}, {Ranc}, {Rattenbury}, {Satoh},
  {Shoji}, {Suematsu}, {Sullivan}, {Sumi}, {Suzuki}, {Tristram}, {Yamakawa},
  {Yamawaki}, {Yonehara}, \& {MOA Collaboration}}]{han2019a}
{Han}, C., {Lee}, C.-U., {Udalski}, A., {et~al.} 2020{\natexlab{a}}, \aj, 159,
  134

\bibitem[{{Han} {et~al.}(2020{\natexlab{b}}){Han}, {Udalski}, {Gould},
  {Albrow}, {Chung}, {Hwang}, {Jung}, {Lee}, {Ryu}, {Shin}, {Shvartzvald},
  {Yee}, {Zang}, {Cha}, {Kim}, {Kim}, {Kim}, {Lee}, {Lee}, {Park}, {Pogge},
  {Jee}, {Kim}, {Kim}, {Kim}, {Mr{\'o}z}, {Szyma{\'n}ski}, {Skowron},
  {Poleski}, {Soszy{\'n}ski}, {Pietrukowicz}, {Koz{\l}owski}, \&
  {Ulaczyk}}]{han2020}
{Han}, C., {Udalski}, A., {Gould}, A., {et~al.} 2020{\natexlab{b}}, \aj, 159,
  91

\bibitem[{{Houdashelt} {et~al.}(2000){Houdashelt}, {Bell}, \&
  {Sweigart}}]{houdashelt2000}
{Houdashelt}, M.~L., {Bell}, R.~A., \& {Sweigart}, A.~V. 2000, \aj, 119, 1448

\bibitem[{{Huang} {et~al.}(2018){Huang}, {Andrews}, {Dullemond}, {Isella},
  {P{\'e}rez}, {Guzm{\'a}n}, {{\"O}berg}, {Zhu}, {Zhang}, {Bai}, {Benisty},
  {Birnstiel}, {Carpenter}, {Hughes}, {Ricci}, {Weaver}, \&
  {Wilner}}]{huang2018}
{Huang}, J., {Andrews}, S.~M., {Dullemond}, C.~P., {et~al.} 2018, \apjl, 869,
  L42

\bibitem[{{Hwang} {et~al.}(2019){Hwang}, {Ryu}, {Kim}, {Albrow}, {Chung},
  {Gould}, {Han}, {Jung}, {Shin}, {Shvartzvald}, {Yee}, {Zang}, {Cha}, {Kim},
  {Kim}, {Lee}, {Lee}, {Lee}, {Park}, \& {Pogge}}]{hwang2019}
{Hwang}, K.-H., {Ryu}, Y.-H., {Kim}, H.-W., {et~al.} 2019, \aj, 157, 23

\bibitem[{{Kim} {et~al.}(2016){Kim}, {Lee}, {Park}, {Kim}, {Cha}, {Lee}, {Han},
  {Chun}, \& {Yuk}}]{kim2016}
{Kim}, S.-L., {Lee}, C.-U., {Park}, B.-G., {et~al.} 2016, \jkas, 49, 37

\bibitem[{{Mr{\'o}z} {et~al.}(2017){Mr{\'o}z}, {Udalski}, {Skowron}, {Poleski},
  {Koz{\l}owski}, {Szyma{\'n}ski}, {Soszy{\'n}ski}, {Wyrzykowski},
  {Pietrukowicz}, {Ulaczyk}, {Skowron}, \& {Pawlak}}]{mroz2017}
{Mr{\'o}z}, P., {Udalski}, A., {Skowron}, J., {et~al.} 2017, \nat, 548, 183

\bibitem[{{Mr{\'o}z} {et~al.}(2018){Mr{\'o}z}, {Ryu}, {Skowron}, {Udalski},
  {Gould}, {Szyma{\'n}ski}, {Soszy{\'n}ski}, {Poleski}, {Pietrukowicz},
  {Koz{\l}owski}, {Pawlak}, {Ulaczyk}, {OGLE Collaboration}, {Albrow}, {Chung},
  {Jung}, {Han}, {Hwang}, {Shin}, {Yee}, {Zhu}, {Cha}, {Kim}, {Kim}, {Kim},
  {Lee}, {Lee}, {Lee}, {Park}, {Pogge}, \& {KMTNet Collaboration}}]{mroz2018}
{Mr{\'o}z}, P., {Ryu}, Y.-H., {Skowron}, J., {et~al.} 2018, \aj, 155, 121

\bibitem[{{Mr{\'o}z} {et~al.}(2019{\natexlab{a}}){Mr{\'o}z}, {Udalski},
  {Skowron}, {Szyma{\'n}ski}, {Soszy{\'n}ski}, {Wyrzykowski}, {Pietrukowicz},
  {Koz{\l}owski}, {Poleski}, {Ulaczyk}, {Rybicki}, \& {Iwanek}}]{mroz2019b}
{Mr{\'o}z}, P., {Udalski}, A., {Skowron}, J., {et~al.} 2019{\natexlab{a}},
  \apjs, 244, 29

\bibitem[{{Mr{\'o}z} {et~al.}(2019{\natexlab{b}}){Mr{\'o}z}, {Udalski},
  {Bennett}, {Ryu}, {Sumi}, {Shvartzvald}, {Skowron}, {Poleski},
  {Pietrukowicz}, {Koz{\l}owski}, {Szyma{\'n}ski}, {Wyrzykowski},
  {Soszy{\'n}ski}, {Ulaczyk}, {Rybicki}, {Iwanek}, {KMTNet Collaboration},
  {Albrow}, {Chung}, {Gould}, {Han}, {Hwang}, {Jung}, {Shin}, {Yee}, {Zang},
  {Cha}, {Kim}, {Kim}, {Kim}, {Lee}, {Lee}, {Lee}, {Park}, {Pogge}, {MOA
  Collaboration}, {Abe}, {Barry}, {Bhattacharya}, {Bond}, {Donachie}, {Fukui},
  {Hirao}, {Itow}, {Kawasaki}, {Kondo}, {Koshimoto}, {Li}, {Matsubara},
  {Muraki}, {Miyazaki}, {Nagakane}, {Ranc}, {Rattenbury}, {Suematsu},
  {Sullivan}, {Suzuki}, {Tristram}, {Yonehara}, {Wise Group}, {Maoz}, {Kaspi},
  \& {Friedmann}}]{mroz2019}
{Mr{\'o}z}, P., {Udalski}, A., {Bennett}, D.~P., {et~al.} 2019{\natexlab{b}},
  \aap, 622, A201

\bibitem[{{Nataf} {et~al.}(2013){Nataf}, {Gould}, {Fouqu{\'e}}, {Gonzalez},
  {Johnson}, {Skowron}, {Udalski}, {Szyma{\'n}ski}, {Kubiak},
  {Pietrzy{\'n}ski}, {Soszy{\'n}ski}, {Ulaczyk}, {Wyrzykowski}, \&
  {Poleski}}]{nataf2013}
{Nataf}, D.~M., {Gould}, A., {Fouqu{\'e}}, P., {et~al.} 2013, \apj, 769, 88

\bibitem[{{Nemiroff} \& {Wickramasinghe}(1994)}]{nemiroff1994}
{Nemiroff}, R.~J., \& {Wickramasinghe}, W.~A.~D.~T. 1994, \apjl, 424, L21

\bibitem[{{Nielsen} {et~al.}(2019){Nielsen}, {De Rosa}, {Macintosh}, {Wang},
  {Ruffio}, {Chiang}, {Marley}, {Saumon}, {Savransky}, {Ammons}, {Bailey},
  {Barman}, {Blain}, {Bulger}, {Burrows}, {Chilcote}, {Cotten}, {Czekala},
  {Doyon}, {Duch{\^e}ne}, {Esposito}, {Fabrycky}, {Fitzgerald}, {Follette},
  {Fortney}, {Gerard}, {Goodsell}, {Graham}, {Greenbaum}, {Hibon}, {Hinkley},
  {Hirsch}, {Hom}, {Hung}, {Dawson}, {Ingraham}, {Kalas}, {Konopacky},
  {Larkin}, {Lee}, {Lin}, {Maire}, {Marchis}, {Marois}, {Metchev},
  {Millar-Blanchaer}, {Morzinski}, {Oppenheimer}, {Palmer}, {Patience},
  {Perrin}, {Poyneer}, {Pueyo}, {Rafikov}, {Rajan}, {Rameau}, {Rantakyr{\"o}},
  {Ren}, {Schneider}, {Sivaramakrishnan}, {Song}, {Soummer}, {Tallis},
  {Thomas}, {Ward-Duong}, \& {Wolff}}]{nielsen2019}
{Nielsen}, E.~L., {De Rosa}, R.~J., {Macintosh}, B., {et~al.} 2019, \aj, 158,
  13

\bibitem[{{Okuzumi} {et~al.}(2016){Okuzumi}, {Momose}, {Sirono}, {Kobayashi},
  \& {Tanaka}}]{okuzumi2016}
{Okuzumi}, S., {Momose}, M., {Sirono}, S.-i., {Kobayashi}, H., \& {Tanaka}, H.
  2016, \apj, 821, 82

\bibitem[{{Pogson}(1856)}]{pogson}
{Pogson}, N. 1856, \mnras, 17, 12

\bibitem[{{Poleski} \& {Yee}(2019)}]{poleski_yee2019}
{Poleski}, R., \& {Yee}, J.~C. 2019, Astronomy and Computing, 26, 35

\bibitem[{{Poleski} {et~al.}(2014){Poleski}, {Udalski}, {Dong},
  {Szyma{\'n}ski}, {Soszy{\'n}ski}, {Kubiak}, {Pietrzy{\'n}ski},
  {Koz{\l}owski}, {Pietrukowicz}, {Ulaczyk}, {Skowron}, {Wyrzykowski}, \&
  {Gould}}]{poleski2014}
{Poleski}, R., {Udalski}, A., {Dong}, S., {et~al.} 2014, \apj, 782, 47

\bibitem[{{Poleski} {et~al.}(2018){Poleski}, {Gaudi}, {Udalski},
  {Szyma{\'n}ski}, {Soszy{\'n}ski}, {Pietrukowicz}, {Koz{\l}owski}, {Skowron},
  {Wyrzykowski}, \& {Ulaczyk}}]{poleski2018b}
{Poleski}, R., {Gaudi}, B.~S., {Udalski}, A., {et~al.} 2018, \aj, 156, 104

\bibitem[{{Skowron} {et~al.}(2016){Skowron}, {Udalski}, {Koz{\l}owski},
  {Szyma{\'n}ski}, {Mr{\'o}z}, {Wyrzykowski}, {Poleski}, {Pietrukowicz},
  {Ulaczyk}, {Pawlak}, \& {Soszy{\'n}ski}}]{skowron2016}
{Skowron}, J., {Udalski}, A., {Koz{\l}owski}, S., {et~al.} 2016, \actaa, 66, 1

\bibitem[{{Smith} {et~al.}(2007){Smith}, {Wo{\'z}niak}, {Mao}, \&
  {Sumi}}]{smith2007}
{Smith}, M.~C., {Wo{\'z}niak}, P., {Mao}, S., \& {Sumi}, T. 2007, \mnras, 380,
  805

\bibitem[{{Tomaney} \& {Crotts}(1996)}]{tomaney1996}
{Tomaney}, A.~B., \& {Crotts}, A. P.~S. 1996, \aj, 112, 2872

\bibitem[{{Udalski}(2003)}]{udalski2003}
{Udalski}, A. 2003, \actaa, 53, 291

\bibitem[{{Udalski} {et~al.}(2015){Udalski}, {Szyma{\'n}ski}, \&
  {Szyma{\'n}ski}}]{udalski2015}
{Udalski}, A., {Szyma{\'n}ski}, M.~K., \& {Szyma{\'n}ski}, G. 2015, \actaa, 65,
  1

\bibitem[{{Witt} \& {Mao}(1994)}]{witt1994}
{Witt}, H.~J., \& {Mao}, S. 1994, \apj, 430, 505

\bibitem[{{Wo{\'z}niak}(2000)}]{wozniak2000}
{Wo{\'z}niak}, P.~R. 2000, \actaa, 50, 421

\bibitem[{{Wyrzykowski} {et~al.}(2015){Wyrzykowski}, {Rynkiewicz}, {Skowron},
  {Koz{\l}owski}, {Udalski}, {Szyma{\'n}ski}, {Kubiak}, {Soszy{\'n}ski},
  {Pietrzy{\'n}ski}, {Poleski}, {Pietrukowicz}, \& {Pawlak}}]{wyrzykowski2015}
{Wyrzykowski}, {\L}., {Rynkiewicz}, A.~E., {Skowron}, J., {et~al.} 2015, \apjs,
  216, 12

\bibitem[{{Yee} {et~al.}(2012){Yee}, {Shvartzvald}, {Gal-Yam}, {Bond},
  {Udalski}, {Koz{\l}owski}, {Han}, {Gould}, {Skowron}, {Suzuki}, {Abe},
  {Bennett}, {Botzler}, {Chote}, {Freeman}, {Fukui}, {Furusawa}, {Itow},
  {Kobara}, {Ling}, {Masuda}, {Matsubara}, {Miyake}, {Muraki}, {Ohmori},
  {Ohnishi}, {Rattenbury}, {Saito}, {Sullivan}, {Sumi}, {Suzuki}, {Sweatman},
  {Takino}, {Tristram}, {Wada}, {MOA Collaboration}, {Szyma{\'n}ski}, {Kubiak},
  {Pietrzy{\'n}ski}, {Soszy{\'n}ski}, {Poleski}, {Ulaczyk}, {Wyrzykowski},
  {Pietrukowicz}, {OGLE Collaboration}, {Allen}, {Almeida}, {Batista}, {Bos},
  {Christie}, {DePoy}, {Dong}, {Drummond}, {Finkelman}, {Gaudi}, {Gorbikov},
  {Henderson}, {Higgins}, {Jablonski}, {Kaspi}, {Manulis}, {Maoz}, {McCormick},
  {McGregor}, {Monard}, {Moorhouse}, {Mu{\~n}oz}, {Natusch}, {Ngan}, {Ofek},
  {Pogge}, {Santallo}, {Tan}, {Thornley}, {Shin}, {Choi}, {Park}, {Lee}, {Koo},
  \& {{$\mu$}FUN Collaboration}}]{yee2012}
{Yee}, J.~C., {Shvartzvald}, Y., {Gal-Yam}, A., {et~al.} 2012, \apj, 755, 102

\bibitem[{{Yoo} {et~al.}(2004){Yoo}, {DePoy}, {Gal-Yam}, {Gaudi}, {Gould},
  {Han}, {Lipkin}, {Maoz}, {Ofek}, {Park}, {Pogge}, {Mu-Fun Collaboration},
  {Udalski}, {Soszy{\'n}ski}, {Wyrzykowski}, {Kubiak}, {Szyma{\'n}ski},
  {Pietrzy{\'n}ski}, {Szewczyk}, {{\.Z}ebru{\'n}}, \& {OGLE
  Collaboration}}]{yoo2004}
{Yoo}, J., {DePoy}, D.~L., {Gal-Yam}, A., {et~al.} 2004, \apj, 603, 139

\bibitem[{{Zhang} {et~al.}(2015){Zhang}, {Blake}, \& {Bergin}}]{zhang2015}
{Zhang}, K., {Blake}, G.~A., \& {Bergin}, E.~A. 2015, \apjl, 806, L7

\bibitem[{{Zhang} {et~al.}(2018){Zhang}, {Zhu}, {Huang}, {Guzm{\'a}n},
  {Andrews}, {Birnstiel}, {Dullemond}, {Carpenter}, {Isella}, {P{\'e}rez},
  {Benisty}, {Wilner}, {Baruteau}, {Bai}, \& {Ricci}}]{zhang2018}
{Zhang}, S., {Zhu}, Z., {Huang}, J., {et~al.} 2018, \apjl, 869, L47

\end{thebibliography}

\end{document}